# Nonequivalence of expected free energy and revised system integrated information in noisy permutation networks

Katsuaki Tanabe*
Department of Chemical Science and Engineering, Kyoto University, Nishikyo, Kyoto 615-8510, Japan
School of Chemical Science and Technology, Kyoto University, Nishikyo, Kyoto 615-8510, Japan
Kyoto MPI Inc., Sakyo, Kyoto 606-8501, Japan
*Contact author: tanabe@cheme.kyoto-u.ac.jp

ABSTRACT. The free-energy principle (FEP) and active inference describe adaptive systems through probabilistic generative models and variational or expected free-energy functionals, whereas integrated information theory (IIT) characterizes the intrinsic cause–effect organization of physical systems. Here we separate two questions that are often conflated: whether revised IIT quantities admit local free-energy-like representations and whether a scalar expected free energy can determine system integrated information. First, we show that the 2026 IIT intrinsic-differentiation and specification terms admit exact representations in surprisal coordinates, and hence in optimized variational-free-energy coordinates under exact inference. We then construct noisy binary permutation networks in which causal topology varies while every local transition reliability is fixed. In a homogeneous current state, every permutation produces exactly the same complete one-step predictive distribution and therefore the same state-risk expected free energy under common preferences and observation models. Revised IIT nevertheless distinguishes the topologies: multicycle permutations have zero whole-system integration because they lack strong connectivity, whereas a single noisy cycle has positive integration at intermediate reliability. For a noisy single cycle of $n$ units with local transition reliability $\eta$, we derive the exact closed-form expression for revised system integrated information, $\varphi_s^{(n)}(\eta) = \min\{2\eta^n\log_2(2\eta), -n\log_2\eta\}$, prove that the directed-set minimum-information partition cuts exactly two actual causal edges, and show that, in the large-system limit, the optimal local reliability approaches unity while the joint specified-state probability approaches one half and the maximum revised system integrated information approaches one ibit. Independent numerical IIT calculations for $n$ = 2–5 and exhaustive directed-partition enumeration through $n$ = 8 reproduce the analytical results to floating-point precision. Thus, local surprisal/free-energy correspondence coexists with a global non-equivalence: equal expected free energy does not imply equal revised IIT system integrated information. Any formal synthesis between IIT and the FEP/active-inference framework must therefore preserve counterfactual causal information beyond a scalar free-energy value alone.

KEYWORDS: Consciousness; Cognition; Integrated information theory; Free energy principle; Active inference

## I. INTRODUCTION

The free-energy principle (FEP) and its associated active-inference framework provide an influential account of adaptive systems, whereas integrated information theory (IIT) approaches consciousness from the standpoint of intrinsic causal organization. IIT grew from the proposal that conscious experience requires both differentiation and integration [1] and was progressively formalized in IIT 3.0 and IIT 4.0 through increasingly explicit causal constructions [2–6]. The FEP and active inference instead describe self-organization, inference, and action using probabilistic generative models and variational or expected free-energy functionals [7–10].

Despite their different origins, the frameworks share a substantial information-theoretic vocabulary. Related statistical-physical approaches have represented integration and stochastic interaction using information geometry [11,12]. Probabilities, surprisal, Kullback–Leibler (KL) divergences, entropy-like quantities, and causal or predictive dependence all enter naturally into their mathematical descriptions [13–16]. Thermodynamic and information-geometric formulations provide further related perspectives [17,18]. This common vocabulary makes a formal comparison natural, but it also creates a risk: local algebraic similarity between formulas need not imply equivalence of the physical quantities or causal structures they evaluate.

The possible relation between IIT and the FEP has therefore become an active topic. Safron proposed an integrated world-modeling framework combining IIT, global neuronal workspace theory (GNWT), the FEP, and active inference [19]. Olesen et al. found empirical covariation between surprisal and IIT-based quantities and presented the result as a preliminary step toward synthesis [20]. More recent work has examined living neuronal cultures [21], maximum-caliber constructions [22], and integrated-information measures in active-inference agents [23]. An adversarial collaborative review has sharpened conceptual distinctions between integrated-information and predictive-processing theories [24], while a large adversarial experiment directly tested IIT and GNWT [25]. Empirical applications include resting-state fMRI [26], in vitro neural networks [27], and broader proposals for Bayesian learning and biological intelligence [28].

These questions also sit within a broader statistical-physics literature on consciousness and information processing [29–31], critical neural dynamics [32], Fisher-information coding [33], and predictive or directed measures of causal dependence [34,35]. The relevant issue is therefore not only whether two scalar quantities correlate in a particular simulation but whether one quantity mathematically determines the other.

The present work approaches this issue in two stages. First, we ask how far the local mathematical analogy can be taken. The refinement of IIT 4.0 in 2026 introduced intrinsic differentiation as a state-specific surprisal [6]. Together with intrinsic specification and partition irreducibility, this leads to exact expressions in negative-log-probability coordinates. Under exact Bayesian inference, such surprisals can in turn be represented as optimized variational free energies. We therefore obtain a genuine local correspondence rather than a merely qualitative analogy.

Second, we ask the stronger system-level question: can a scalar expected free energy determine revised IIT system integrated information? To answer it, we construct noisy permutation networks in which causal wiring is changed while all local transition reliabilities are kept fixed. We deliberately evaluate a homogeneous current state because it is invariant under arbitrary permutations. Consequently, distinct causal architectures can generate exactly the same complete one-step predictive distribution. This makes it possible to separate intact predictive equivalence from counterfactual causal equivalence.

The main results of the present study are threefold. First, we prove that a single-cycle and a multicycle permutation can have identical state-conditioned predictive distributions and identical state-risk expected free energies while possessing different revised IIT system integrated information. This rules out a universal scalar mapping from expected free energy $G$ to revised IIT system integrated information $\varphi_s$ over the noisy-permutation family. Second, for a noisy single $n$-cycle we derive the exact expression $\varphi_s^{(n)}(\eta) = \min\{2\eta^n\log_2(2\eta), -n\log_2\eta\}$. Third, we prove the minimum-information partition (MIP) structure, derive the unique optimal reliability, obtain its large-$n$ asymptotics, and independently verify the results using the current PyPhi implementation and exhaustive partition enumeration. The overall message is therefore not that IIT and free-energy descriptions are unrelated, but that an exact local correspondence coexists with a global causal non-equivalence.

## II. THEORETICAL BACKGROUND

### A. Revised integrated information theory

IIT 4.0 characterizes a candidate physical system $S$ in its current state $s$ through its intrinsic cause–effect power [4,5]. Cause and effect repertoires are constructed interventionally and intrinsically, using causal marginalization rather than an externally imposed observational distribution. The 2026 refinement adds an explicit tradeoff between intrinsic differentiation and intrinsic specification [6].

For a selected effect state $s'_e$, let $p = p_e(s'_e|s)$ denote its intact conditional probability and $q = p_e(s'_e)$ its unconstrained probability. Intrinsic effect differentiation is the state-specific surprisal

$$i_{\text{diff}}^e(s,s'_e) = -\log_2 p. \tag{1}$$

Intrinsic specification measures the extent to which the intact system selectively increases the probability of the specified effect state relative to its unconstrained probability,

$$i_{\text{spec}}^e(s,s'_e) = p\log_2(p/q). \tag{2}$$

The revised intrinsic effect information is limited by the weaker of differentiation and specification,

$$ii_e(s) = \min\{i_{\text{diff}}^e(s),\ i_{\text{spec}}^e(s)\}. \tag{3}$$

Analogous quantities are defined on the cause side, and the overall system intrinsic information is

$$ii(s) = \min\{ii_c(s),\ ii_e(s)\}. \tag{4}$$

For a directed system partition $\theta$, severed causal influences are replaced by causally marginalized inputs. If $p_e^\theta(s'_e|s)$ is the partitioned probability of the selected effect state, the corresponding effect irreducibility is

$$\varphi_e(s,\theta) = p_e(s'_e|s)\log_2[p_e(s'_e|s)/p_e^\theta(s'_e|s)]. \tag{5}$$

Cause irreducibility is defined analogously. The system MIP is selected by minimizing normalized integrated information, with the normalization accounting for the maximum number of connections that the partition can affect. Once the MIP is identified, the absolute unnormalized cause and effect irreducibilities associated with that partition enter the final system measure. Under the 2026 refinement,

$$\varphi_s(s) = \min\{\varphi_c(s),\ \varphi_e(s),\ ii(s)\}. \tag{6}$$

This final minimum is essential for the results below. A deterministic system may remain highly sensitive to causal partition while having zero intrinsic differentiation because the selected state is certain. Hence the revised $\varphi_s$ can vanish at determinism even when partition-based cause and effect irreducibilities are positive. Strong connectivity is also required for positive whole-system integration [4,5]. Following IIT convention, intrinsic-information values are reported in intrinsic bits (ibits); an ibit is a probability-weighted pointwise information value expressed using base-2 logarithms [5,6].

### B. Variational free energy and expected free energy

Variational free energy (VFE) and expected free energy (EFE) are distinct quantities and are used for different purposes in the FEP and active-inference literature [7–10]. Let $o$ denote observed data, $z$ latent variables, $P(o,z)$ a generative model, and $Q(z)$ a variational distribution approximating the posterior $P(z|o)$; $E_Q[\cdot]$ denotes expectation with respect to $Q$. VFE is

$$F[Q] = E_Q[\ln Q(z) - \ln P(o,z)]. \tag{7}$$

Using Bayes' rule, this can be rewritten as

$$F[Q] = D_{\mathrm{KL}}[Q(z) \parallel P(z|o)] - \ln P(o). \tag{8}$$

Because the KL divergence is nonnegative, $F[Q] \geq -\ln P(o)$, and under exact inference $Q(z) = P(z|o)$,

$$F_{\min} = -\ln P(o). \tag{9}$$

Equation (9) provides the precise sense in which a surprisal can be represented as an optimized variational free energy. This correspondence does not imply that every negative log-probability appearing in IIT is itself the VFE of an active-inference agent.

EFE is a prospective quantity used to evaluate predicted futures. Several related formulations coexist, and state-risk and outcome-risk forms should not be treated as automatically interchangeable [8,10]. For the principal counterexample we adopt a state-risk-plus-ambiguity construction. For future hidden state $S'$, future observation $O'$, predicted future-state distribution $Q(S')$, and a fixed one-step transition model, the expected free energy is

$$G = D_{\mathrm{KL}}[Q(S') \parallel C_S(S')] + E_Q\, H[P(O'|S')], \tag{10}$$

where $C_S(S')$ denotes prior preferences over future states and $H[\cdot]$ denotes Shannon entropy. Because no policy optimization is considered here, explicit policy conditioning is suppressed throughout. In the fully observed special case $P(O'|S') = \delta_{O',S'}$, where $\delta$ is the Kronecker delta, the ambiguity term vanishes and

$$G = D_{\mathrm{KL}}[Q(S') \parallel C_S(S')]. \tag{11}$$

Active-inference expressions are conventionally written using natural logarithms. When numerical values are compared with IIT quantities, we use $G_2 = G/\ln 2$ so that EFE is expressed in bits. The logarithm base does not affect equality, ordering, minimization, or the non-equivalence results below.

### C. Scope of the comparison

The analysis does not attempt to identify the FEP as a whole with IIT, nor does it claim that an active-inference generative model cannot explicitly represent causal topology. The narrower question is whether revised IIT system integrated information can be universally determined from a scalar free-energy quantity. To disprove a universal mapping $\varphi_s = f(G)$, it is sufficient to construct two systems having the same $G$ but different $\varphi_s$. Before constructing that counterexample, however, it is useful to make the local mathematical correspondence explicit.

## III. REVISED IIT IN SURPRISAL AND FREE-ENERGY COORDINATES

### A. Effect-side coordinates

Consider a system in state s and an effect state $s'_e$ selected by IIT. Define the intact, unconstrained, and partitioned probabilities

$$p \equiv p_e(s'_e|s), \quad q \equiv p_e(s'_e), \quad r_\theta \equiv p_e^{\theta}(s'_e|s). \tag{12}$$

Introduce the base-2 surprisal coordinates

$$F_I = -\log_2 p, \quad F_U = -\log_2 q, \quad F_P^{\theta} = -\log_2 r_\theta. \tag{13}$$

Intrinsic differentiation is immediately

$$i_{\mathrm{diff}}^{e} = F_I. \tag{14}$$

Intrinsic specification becomes

$$i_{\mathrm{spec}}^{e} = p(F_U - F_I). \tag{15}$$

Likewise, the partition-dependent effect irreducibility can be written

$$\varphi_e(\theta) = p(F_P^{\theta} - F_I). \tag{16}$$

Thus, three central effect-side IIT quantities are exact combinations of the same surprisal coordinates. Revised IIT simultaneously asks how non-deterministic the intact repertoire is, how selectively the intact system specifies a state relative to an unconstrained baseline, and how much of that specification is lost under causal partition.

### B. Cause-side coordinates

The cause side requires slightly more care because IIT separates selectivity for a candidate cause state from the forward likelihood entering the intrinsic-difference ratio. Let $\alpha = p_c(s'_c|s)$ denote cause-side selectivity and define the intact, unconstrained, and partitioned forward likelihoods $\ell_I = p_c(s|s'_c)$, $\ell_U = p_c(s)$, and $\ell_P^{\theta} = p_c^{\theta}(s|s'_c)$. Introduce

$$F_\alpha = -\log_2 \alpha, \quad L_I = -\log_2 \ell_I, \; L_U = -\log_2 \ell_U, \; L_P^{\theta} = -\log_2 \ell_P^{\theta}. \tag{17}$$

Then

$$i_{\text{diff}}{}^{c} = F_{\alpha}, \quad (18)$$
$$i_{\text{spec}}{}^{c} = \alpha(L_U - L_I),$$
$$\varphi_c(\theta) = \alpha(L_P{}^{\theta} - L_I).$$

For the symmetric noisy permutation channels considered below, the corresponding cause and effect quantities coincide in the homogeneous states used for the analytical examples.

### C. What the local correspondence does and does not establish

Equations (14)–(18) establish exact algebraic correspondences between revised IIT quantities and negative-log-probability coordinates. Under exact Bayesian inference, each individual surprisal coordinate can be represented formally as an optimized variational free-energy coordinate through Eq. (9). These are genuine local correspondences.

They do not imply global equivalence between revised IIT and the FEP or active inference. First, the IIT expressions depend on different probability models: intact, unconstrained, and causally partitioned repertoires. Second, partitioned probabilities are generated by causal interventions and marginalization, not merely by the intact predictive model. Third, $\varphi_s$ is obtained through an optimization over causal partitions followed by a minimum across several existence requirements. A single intact surprisal or EFE value therefore need not identify the MIP, the partitioned probabilities, or the system's strong connectivity. The counterexample below isolates precisely this missing causal information.

## IV. NOISY PERMUTATION NETWORKS

### A. Model definition

Consider $n$ binary variables $X = (X_1, \ldots, X_n)$ with one-step dynamics indexed by a permutation $\sigma \in S_n$, where $S_n$ is the symmetric group. The subsequent variables are

$$Y_i = X_{\sigma(i)} \oplus N_i, \quad (19)$$
$$P(N_i = 0) = \eta, \;\; P(N_i = 1) = 1-\eta, \; 1/2 \leq \eta \leq 1,$$

where $\oplus$ denotes the XOR operation. The corresponding transition kernel is

$$T_\sigma(y|x) = \textstyle\prod_i \eta^{1[y_i = x_{\sigma(i)}]}(1-\eta)^{1[y_i \neq x_{\sigma(i)}]}, \quad (20)$$

where $1[\cdot]$ is the indicator function. Every node has exactly one incoming and one outgoing causal edge. Changing $\sigma$ therefore changes only the routing of causal influences; $n$, $\eta$, and the local binary channel are held fixed.

### B. Exact predictive equivalence in a homogeneous state

We deliberately evaluate either of the two homogeneous states $x^{(b)} = (b, \ldots, b)$, with $b \in \{0,1\}$. These states are invariant under arbitrary permutation of the units. Because $x^{(b)}_{\sigma(i)} = b$ for every $i$ and every $\sigma$,

$$T_\sigma(y|x^{(b)}) = \textstyle\prod_i \eta^{1[y_i = b]}(1-\eta)^{1[y_i \neq b]}, \quad (21)$$

which contains no dependence on $\sigma$. Let $Q_\sigma(Y|x)$ denote the one-step predictive distribution induced by the transition kernel $T_\sigma(\cdot|x)$. Hence, for any two permutations,

$$Q_{\sigma_1}(Y|x^{(b)}) = Q_{\sigma_2}(Y|x^{(b)}) = Q_\eta(Y). \quad (22)$$

This equality is stronger than equality of a scalar entropy or free energy: the complete one-step predictive distribution is identical across causal architectures. For $b = 1$ and a future state $y$ containing $m(y)$ ones, $Q_\eta(y) = \eta^{m(y)}(1-\eta)^{n-m(y)}$.

### C. Expected-free-energy degeneracy

Embed each physical transition model in the same one-step state-risk active-inference construction. The causal architecture $\sigma$ is treated as a property of the physical transition model, not as a policy. Using the same preference distribution $C_S$ and observation model $P(O'|S')$ for every architecture, Eq. (22) implies that both the state-risk and ambiguity terms in Eq. (10) are identical. Therefore

$$G_{\sigma_1}(x^{(b)}) = G_{\sigma_2}(x^{(b)}). \quad (23)$$

Uniform preferences are not required. In the fully observed special case, $G_\sigma = D_{\text{KL}}[Q_\eta \| C_S]$ for arbitrary common $C_S$. If $C_S$ is uniform over $2^n$ states, then in bits

$$G_2(\eta) = n[1 - h_2(\eta)], \quad (24)$$

where $h_2$ is the binary entropy.

### D. Uniform-ensemble predictive invariance

A complementary construction averages current states uniformly, $Q(X) = U_{2n}$, where $U_{2n}$ is the uniform distribution over the $2^n$ binary states. A permutation of a uniform binary vector remains uniform, and the independent binary-symmetric channels preserve the uniform output. Hence $Q_\sigma(Y) = U_{2n}$ for every $\sigma$. The conditional transition entropy is

$$H_2(Y|X) = nh_2(\eta), \quad (25)$$

and because $H_2(Y) = n$, the mutual information is

$$I(X;Y) = n[1 - h_2(\eta)]. \quad (26)$$

These ensemble equalities are complementary to, but not required for, the state-conditioned non-equivalence theorem. A representative three-unit example, including the identical predictive distributions and the corresponding topology-dependent IIT values derived below, is illustrated in Fig. 1.

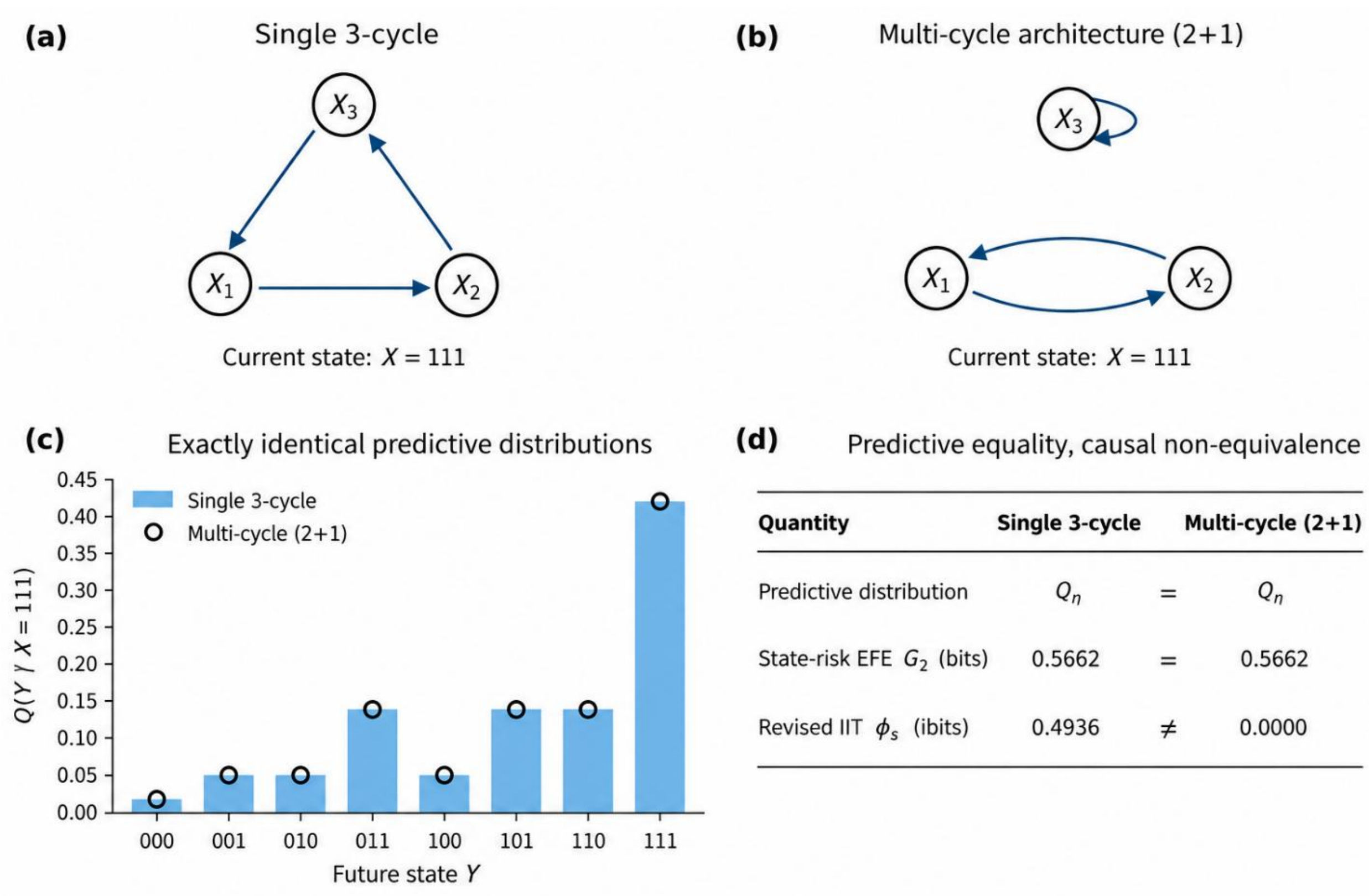

FIG. 1. Predictive equivalence does not imply equivalent intrinsic causal integration. (a) A noisy single 3-cycle and (b) a multicycle permutation consisting of a reciprocal two-node cycle and a self-loop, both evaluated at $X = 111$ and $\eta = 0.75$. (c) The complete one-step predictive distributions coincide exactly for the two architectures. (d) The fully observed state-risk EFE with uniform future-state preferences is also identical, $G_2 \approx 0.5662$ bits, while revised IIT gives $\varphi_s \approx 0.4936$ ibits for the single cycle and $\varphi_s = 0$ for the multicycle system.

## V. PREDICTIVE EQUIVALENCE AND CAUSAL NON-EQUIVALENCE

### A. Cycle decomposition and strong connectivity

Every permutation decomposes uniquely into disjoint directed cycles. Because the causal graph of Eq. (19) has exactly one incoming and one outgoing edge at each node, its strongly connected components are exactly these cycles. The complete $n$-node graph is therefore strongly connected if and only if $\sigma$ is a single $n$-cycle.

A multicycle permutation consequently lacks whole-system strong connectivity and has

$$\varphi_s(\sigma_m) = 0. \tag{27}$$

Individual cycles within the architecture may form integrated subsystems; Eq. (27) concerns the complete $n$-node candidate system.

### B. Positivity for a noisy single cycle

Consider a single $n$-cycle in the homogeneous state $1^n$ with $1/2 < \eta < 1$. The intact all-ones probability is $p = \eta^n$ and the unconstrained all-ones probability is $q = 2^{-n}$. Since $\eta > 1/2$, $p > q$ and intrinsic specification is positive. Since $\eta < 1$, $p < 1$ and intrinsic differentiation is also positive. Any nontrivial directed partition must sever at least one actual cycle edge. A severed binary input is causally marginalized and contributes probability 1/2 rather than $\eta$ to the specified target value. Therefore, every nontrivial admissible partition has positive irreducibility. The finite minimum over partitions is consequently positive, and

$$\varphi_s(\sigma_c) > 0 \text{ for } 1/2 < \eta < 1. \tag{28}$$

### C. Main non-equivalence theorem

Combining the predictive invariance with the topological distinction yields the central result. For every $n \geq 2$ and $1/2 < \eta < 1$, choose a single-cycle permutation $\sigma_c$ and any multicycle permutation $\sigma_m$. In the same homogeneous current state and under the same state-risk generative model,

$$Q_{\sigma c}(Y|1^n) = Q_{\sigma m}(Y|1^n), \tag{29}$$

and hence

$$G_{\sigma c} = G_{\sigma m}. \tag{30}$$

Nevertheless,

$$\varphi_s(\sigma_m) = 0 < \varphi_s(\sigma_c). \quad (31)$$

Therefore, there is no universal scalar function $f$ satisfying

$$\varphi_s = f(G) \quad (32)$$

over the noisy-permutation family. This is a failure of functional identifiability rather than a statement about weak correlation, measurement units, or numerical approximation. Moreover, the failure occurs before scalar compression into $G$ because the complete intact one-step predictive distribution itself is identical. The missing information is counterfactual: which nodes make causal differences to which other nodes and how those differences change under intervention and partition.

## VI. CLOSED-FORM REVISED IIT FOR A NOISY SINGLE CYCLE

### A. Intrinsic information of the intact cycle

For a noisy single $n$-cycle in state $1^n$, every unit remains at the specified value with probability $\eta$, so the intact probability of the all-ones future state is

$$p = \eta^n. \quad (33)$$

Causal marginalization yields a uniform unconstrained distribution for each binary unit, hence

$$q = 2^{-n}. \quad (34)$$

The intrinsic-differentiation and intrinsic-specification terms in 2026 [6] are therefore

$$i_{\text{diff}} = -n\log_2\eta, \quad (35)$$

$$i_{\text{spec}} = n\eta^n\log_2(2\eta). \quad (36)$$

### B. Irreducibility across an arbitrary directed partition

Let $\theta$ be a directed partition that severs exactly $k$ actual causal edges of the cycle. Each unsevered input contributes $\eta$ to the specified all-ones state. Causal marginalization of a severed binary input gives probability $(\eta+(1-\eta))/2 = 1/2$ for the specified target value. Hence

$$p^\theta = \eta^{n-k}2^{-k}. \quad (37)$$

The unnormalized effect irreducibility is

$$\varphi_e(\theta) = k\eta^n\log_2(2\eta). \quad (38)$$

By the symmetry of the noisy permutation channel, the cause-side value is identical,

$$\varphi_c(\theta) = \varphi_e(\theta) = k\eta^n\log_2(2\eta). \quad (39)$$

### C. Reduction of the MIP problem

Let $k_\theta$ denote the number of actual causal edges severed by partition $\theta$, and let $M_\theta$ denote the normalization denominator, equivalently the number of possible directed connections severed by the corresponding directed-set cut matrix. For fixed $n$ and $\eta$, the positive factor $\eta^n\log_2(2\eta)$ is common to all partitions. The normalized MIP search therefore reduces exactly to

$$\theta_{\text{MIP}} = \arg\min_\theta k_\theta/M_\theta. \quad (40)$$

Appendix A provides the complete combinatorial proof. For every $n \geq 3$ the global optimum severs exactly two actual cycle edges and has

$$k_{\text{MIP}} = 2,\ M_{\text{MIP}} = \lfloor 4n^2/7 \rfloor. \quad (41)$$

For $n = 2$, one-edge and two-edge partitions tie after normalization; the IIT tie-resolution procedure selects the two-edge partition because it has larger unnormalized integrated information. Therefore, for every $n \geq 2$,

$$\varphi_c = \varphi_e = 2\eta^n\log_2(2\eta). \quad (42)$$

### D. Closed-form system integrated information

For $n \geq 2$, the intrinsic-specification term $n\eta^n\log_2(2\eta)$ is never smaller than the two-edge MIP term $2\eta^n\log_2(2\eta)$. Hence, the final minimum in the 2026 refinement is determined by the MIP term and intrinsic differentiation. We obtain

$$\varphi_s^{(n)}(\eta) = \min\{2\eta^n\log_2(2\eta),\ -n\log_2\eta\}. \quad (43)$$

This is the principal closed-form result. The partition-limited branch vanishes at complete randomness $\eta = 1/2$, whereas the differentiation branch vanishes at complete determinism $\eta = 1$. Hence

$$\varphi_s^{(n)}(1/2) = \varphi_s^{(n)}(1) = 0. \quad (44)$$

The nonmonotonic reliability dependence is therefore an analytical consequence of the differentiation–partition tradeoff, not an empirical feature fitted from the numerical calculations. The resulting nonmonotonic behavior for the reciprocal two-node system, together with the corresponding uniform-preference EFE, is shown in Fig. 2.

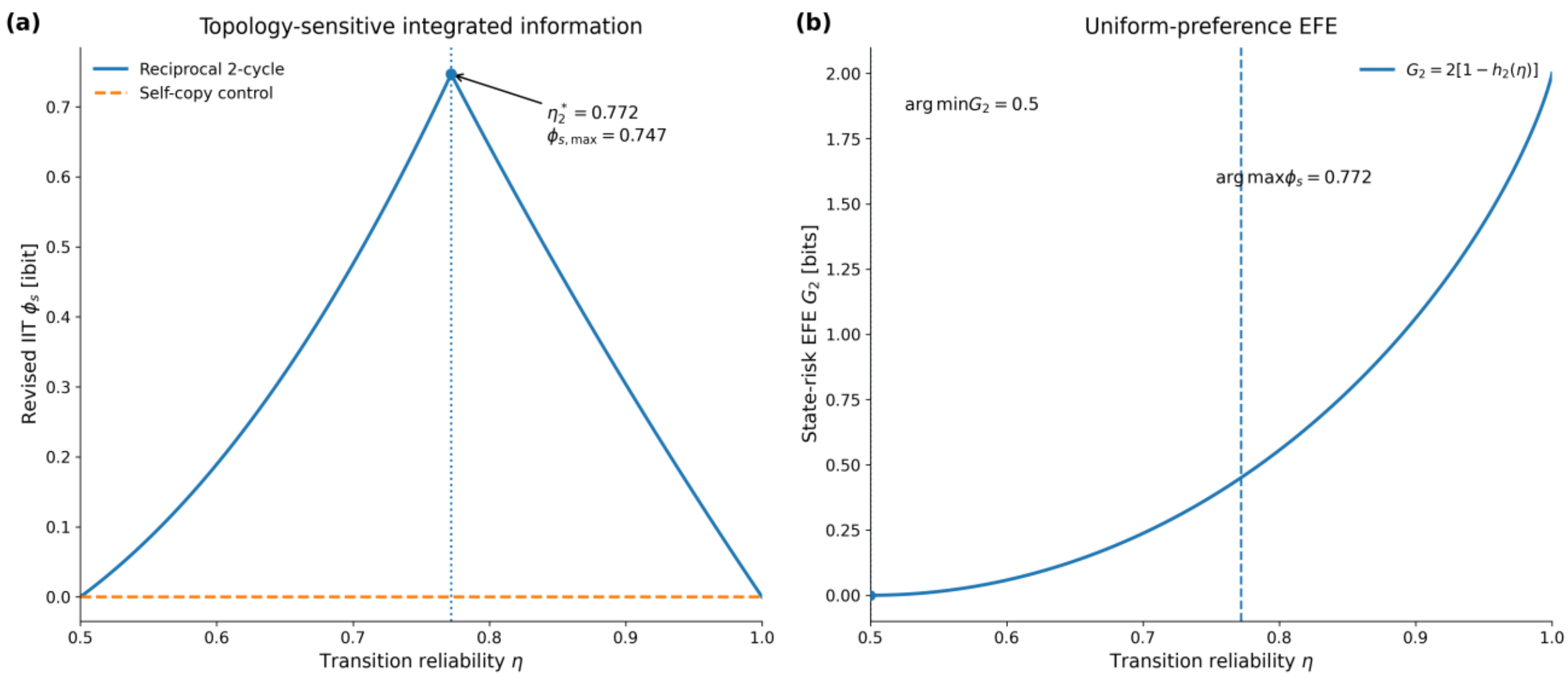

FIG. 2. Exact two-node counterexample and reliability optimization. (a) Revised system integrated information for a reciprocal noisy 2-cycle and a disconnected self-copy control. The reciprocal system follows Eq. (43) with $n = 2$ and reaches its unique maximum at $\eta \approx 0.772$, whereas the self-copy system has $\varphi_s = 0$ throughout because it is not strongly connected. (b) For the fully observed state-risk construction with uniform future-state preferences, $G_2 = 2[1-h_2(\eta)]$ is minimized at $\eta = 1/2$, illustrating an optimization mismatch for this explicit preference choice.

## VII. RELIABILITY OPTIMIZATION AND ASYMPTOTIC CONSEQUENCES

### A. Unique interior maximum

Define the partition-limited and differentiation-limited branches

$$A_n(\eta) = 2\eta^n \log_2(2\eta), \tag{45}$$

$$D_n(\eta) = -n\log_2\eta. \tag{46}$$

For $1/2 < \eta < 1$, $A_n$ is strictly increasing because

$$dA_n/d\eta = [2\eta^{n-1}/\ln 2][n\ln(2\eta)+1] > 0, \tag{47}$$

whereas $D_n$ is strictly decreasing,

$$dD_n/d\eta = -n/(\eta\ln 2) < 0. \tag{48}$$

At $\eta = 1/2$, $A_n = 0$ and $D_n = n$; at $\eta = 1$, $A_n = 2$ and $D_n = 0$. The two branches therefore cross exactly once in the open interval. The unique maximizing reliability $\eta_n^*$ satisfies

$$-n\log_2\eta_n^* = 2(\eta_n^*)^n \log_2(2\eta_n^*). \tag{49}$$

At the crossing, the two limiting requirements are balanced. The first numerical values are summarized in Table I. The reliability dependence of $\varphi_s^{(n)}$ for several cycle sizes is shown in Fig. 3(a), while Fig. 3(b) illustrates explicitly how the partition-limited and differentiation-limited branches determine their lower envelope.

**TABLE I. Optimal reliability and maximum revised integrated information for selected cycle sizes.**

| $n$ | $\eta_n^*$ | $\varphi_{s,max}$ (ibits) |
|---|---|---|
| 2 | 0.771960 | 0.746806 |
| 3 | 0.827089 | 0.821658 |
| 4 | 0.861148 | 0.862666 |
| 5 | 0.884120 | 0.888430 |
| 10 | 0.936756 | 0.942545 |

### B. Finite-size bound below one ibit

A useful reference point is $\eta_0 = 2^{-1/n}$, for which $\eta_0^n = 1/2$. At this reliability $D_n(\eta_0) = 1$, while

$$A_n(\eta_0) = 1-1/n. \tag{50}$$

Thus $A_n(\eta_0) < D_n(\eta_0)$, so the crossing occurs at $\eta_n^* > 2^{-1/n}$. Consequently $-n\log_2\eta_n^* < 1$ and therefore

$$\varphi_{s,\max}^{(n)} < 1 \text{ ibit for finite } n. \tag{51}$$

### C. Expected-free-energy optimization and preference dependence

For the fully observed state-risk construction with uniform future-state preferences,

$$G_2(\eta) = n[1-h_2(\eta)]. \tag{52}$$

Binary entropy is maximal at $\eta = 1/2$, hence

$$\arg\min_\eta G_2 = 1/2. \tag{53}$$

By contrast, $\arg\max_\eta \varphi_s = \eta_n^* > 1/2$. Therefore, the two optima do not coincide under this explicit preference choice. This should not be misread as a universal opposition between EFE and IIT. The EFE optimum is preference dependent. If future-state preferences factorize with preferred reliability $c$, then

$$G_{2,c}(\eta) = n\, D_{KL}^{(2)}[\mathrm{Bern}(\eta) \parallel \mathrm{Bern}(c)], \tag{54}$$

where $D_{KL}^{(2)}$ is the base-2 KL divergence and Bern(·) denotes a Bernoulli distribution. This expression is uniquely minimized at

$$\arg\min_\eta G_{2,c} = c. \tag{55}$$

Thus, one can choose $c = \eta_n^*$ and make the two optima coincide by construction. The relevant conclusion is not universal opposition but the absence of a preference-independent identification.

### D. Large-system limit

Set $a_n = -n\ln\eta_n^*$. Then $\eta_n^* = \exp(-a_n/n)$ and $(\eta_n^*)^n = e^{-a_n}$. Rewriting Eq. (49) in natural logarithms yields

$$a_n = 2e^{-a_n}(\ln 2 - a_n/n). \quad (56)$$

As $n \to \infty$, any limiting value a must satisfy

$$ae^a = 2\ln 2. \quad (57)$$

The positive solution is exactly $a = \ln 2$ because $(\ln 2)e^{\ln 2} = 2\ln 2$. Therefore

$$\eta_n^* \to 1, \quad (\eta_n^*)^n \to 1/2, \quad (58)$$
$$\varphi_{s,\max}^{(n)} \to 1 \text{ ibit}.$$

The first-order corrections follow by writing $a_n = \ln 2 + b/n + O(n^{-2})$, which gives $b = -\ln 2/(1+\ln 2)$. Hence

$$\eta_n^* = 1-(\ln 2)/n+O(n^{-2}), \quad (59)$$

$$\varphi_{s,\max}^{(n)} = 1-1/[(1+\ln 2)n] + O(n^{-2}). \quad (60)$$

The large-system optimum is therefore initially counterintuitive: each local causal transmission becomes increasingly reliable, yet the joint specified-state probability approaches one half. The system retains approximately one ibit of intrinsic differentiation even as every local interaction approaches determinism. The convergence of the optimal reliability and maximum integrated information with increasing system size is shown in Figs. 4(a) and 4(b), respectively.

### E. Combinatorial amplification

The topology difference becomes increasingly pronounced with system size. There are $n!$ permutations of $n$ units, but only $(n-1)!$ single $n$-cycles. Therefore

$$P(\text{single cycle}) = 1/n, \quad (61)$$

$$P(\text{multicycle}) = 1-1/n. \quad (62)$$

At fixed $n$, $\eta$, homogeneous state, preferences, and observation model, all $n!$ permutation architectures have the same EFE. Yet the multicycle fraction has zero whole-system integration. Thus, the fraction of predictively degenerate architectures lacking whole-system $\varphi_s$ approaches unity as $n$ grows. This combinatorial amplification is illustrated in Fig. 4(c).

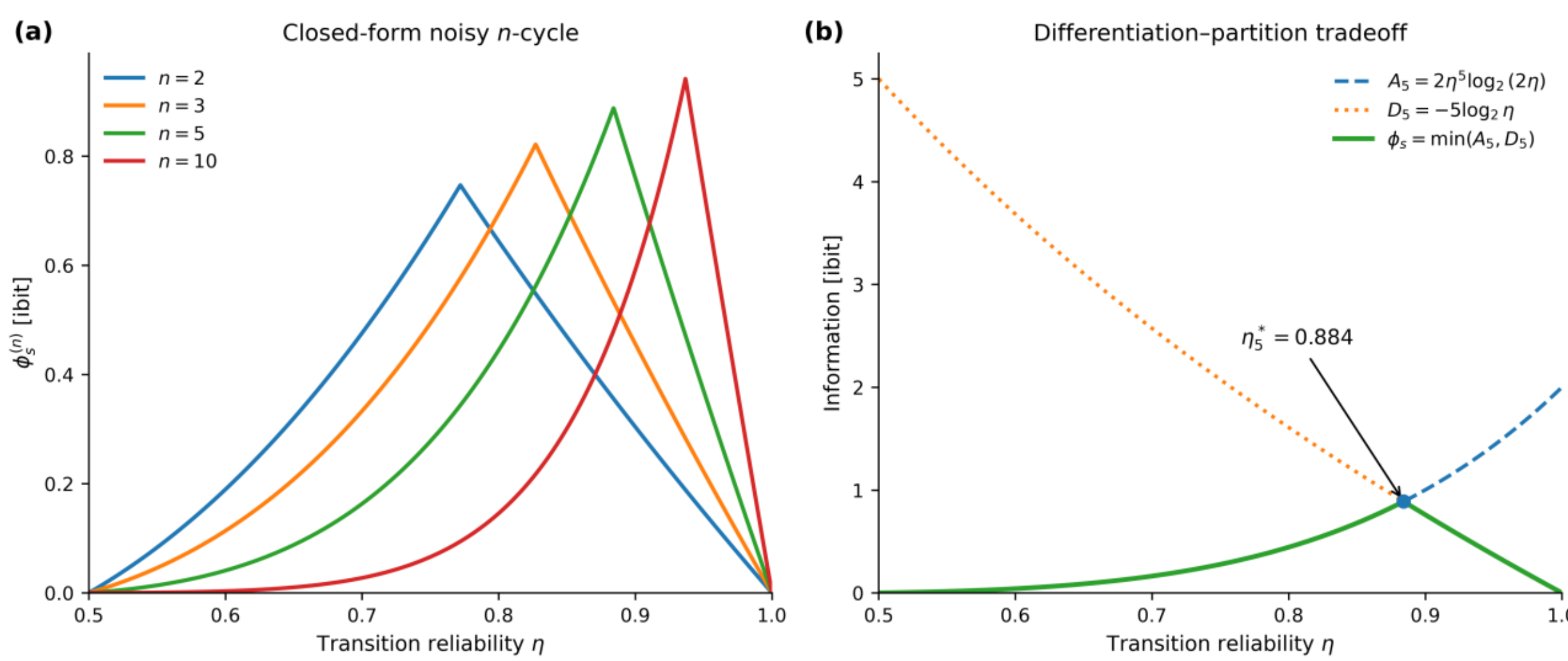

FIG. 3. Closed-form revised integrated information for noisy single cycles. (a) $\varphi_s^{(n)}(\eta)$ for $n = 2, 3, 5$, and 10. Increasing $n$ shifts the maximum toward higher reliability and raises the peak toward one ibit. (b) For $n = 5$, the partition-limited term $A_5 = 2\eta^5\log_2(2\eta)$ increases with $\eta$ while intrinsic differentiation $D_5 = -5\log_2\eta$ decreases; $\varphi_s$ is their lower envelope.

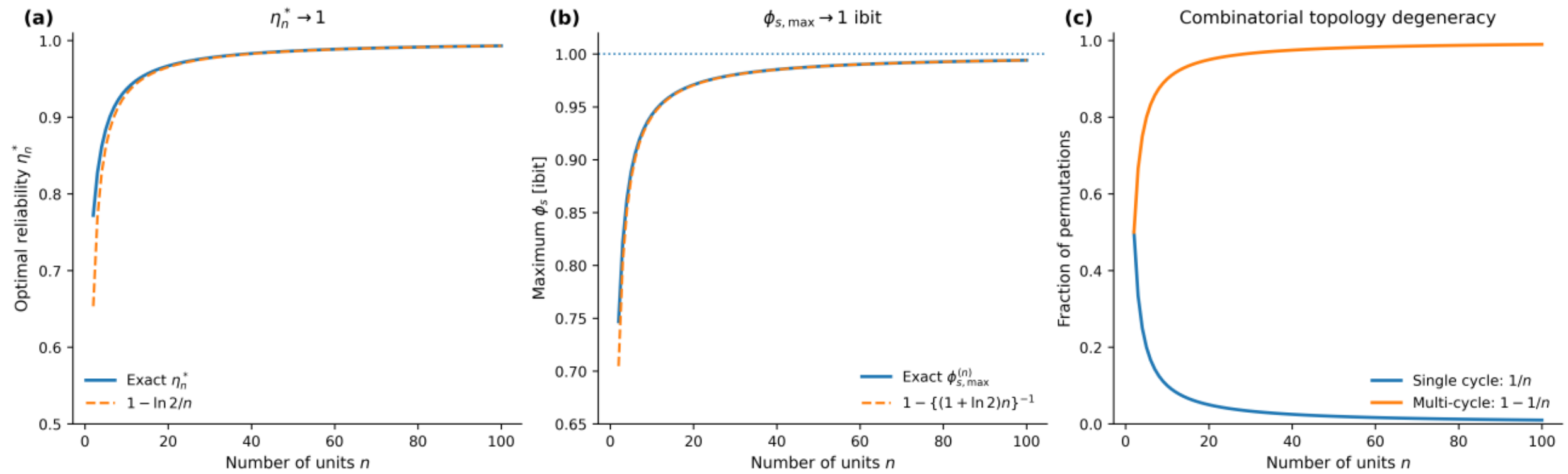

FIG. 4. Large-system scaling and combinatorial amplification. (a) The optimal reliability approaches unity with first-order form $1-\ln 2/n$. (b) The maximum revised system integrated information approaches one ibit as $1-\{(1+\ln 2)n\}^{-1}$. (c) Among $n!$ permutations, the single-cycle fraction is $1/n$ and the multicycle fraction is $1-1/n$.

## VIII. COMPUTATIONAL METHODS

### A. Analytical calculations

All theoretical calculations used the binary Markov dynamics of Eq. (19). Unless otherwise stated, revised IIT quantities were evaluated in the homogeneous all-ones state $X = 1^n$. This state was deliberately chosen because it is invariant under arbitrary permutations and therefore allows causal topology to vary while leaving the complete intact one-step predictive distribution unchanged. All IIT information quantities were expressed in base-2 logarithms. EFE values were converted to bits by division by ln2 when displayed alongside IIT quantities.

The optimum $\eta_n^*$ was obtained numerically only after the analytical crossing equation had been derived. Bisection was used to solve Eq. (49); existence and uniqueness follow analytically from Eqs. (47) and (48). Large-$n$ expansions were obtained by substituting $a_n = -n\ln\eta_n^*$ into the exact crossing equation and expanding in powers of $1/n$.

### B. PyPhi validation

Transition probability matrices were constructed directly from the noisy-copy rule. System analyses were performed using the 2026 IIT 4.0 formalism and directed-set partitions. Multi-cycle controls that lacked strong connectivity returned null system irreducibility analyses because the complete candidate system was not strongly connected. Cause and effect quantities, intrinsic differentiation, and the final system $\varphi_s$ were extracted directly from the current PyPhi system-irreducibility analysis objects.

### C. Independent partition enumeration

The partition audit was implemented separately from the calculation of cause and effect repertoires. For each distinct directed-set cut matrix, $k$ was determined by intersecting the cut matrix with the actual directed edges of the canonical $n$-cycle. $M$ was the total number of directed matrix entries severed by that cut. Because all partitions share the factor $\eta^n\log_2(2\eta)$, minimizing normalized integrated information reduces to minimizing $k/M$. Assertions checked the predicted two-edge optimum and $M = \lfloor 4n^2/7 \rfloor$ for $n = 2, \ldots, 8$.

## IX. COMPUTATIONAL VALIDATION

### A. Implementation and 2026 benchmark

The analytical results were independently checked using the current development implementation of PyPhi [36], which supports IIT 4.0 with the 2026 intrinsic-differentiation refinement. Calculations used PyPhi version 2.1.dev1707+g1f47a1e20, Python 3.13.12, the 2026 IIT 4.0 formalism, causal marginalization, and directed-set system partitions. The complete node set was evaluated as the candidate system in each topology comparison.

As an implementation check, the one-unit differentiation–specification benchmark associated with the 2026 refinement was reproduced. The optimum occurred at $p^* \approx 0.744008805$, where $i_{\mathrm{diff}} \approx 0.426608400$ ibits and $i_{\mathrm{spec}} \approx 0.426608400$ ibits, giving $ii_{\max} \approx 0.426608400$ ibits. This verifies that the installed implementation applies the 2026 intrinsic-information constraint as intended.

### B. Two-unit reciprocal and self-copy systems

For the reciprocal two-node noisy copy system, Eq. (43) specializes to $\varphi_s^{(2)}(\eta) = \min\{2\eta^2\log_2(2\eta), -2\log_2\eta\}$. This expression was evaluated against PyPhi at 101 equally spaced points over $0.5 \le \eta \le 1$. The maximum absolute discrepancy was 0 at the reported floating-point precision. The optimum was $\eta_2^* \approx 0.771959574$ with $\varphi_{s,\max} \approx 0.746805594$ ibits.

The disconnected self-copy control returned $\varphi_s = 0$ at all 101 points because the complete two-node system is not strongly connected. At $\eta = 0.75$, the reciprocal system had $\varphi_s = 0.6580828133$, while the self-copy control had $\varphi_s = 0$, despite identical homogeneous-state predictive distributions. Their fully observed uniform-preference state-risk EFE values were also identical, $G_2 = 0.3774437511$ bits.

### C. Exhaustive enumeration of three-unit permutation architectures

All 3! = 6 three-unit permutation architectures were evaluated at $\eta = 0.75$. The set comprises one identity permutation with cycle type 1 + 1 + 1, three permutations with cycle type 2 + 1, and two single 3-cycles. All four multicycle architectures returned $\varphi_s = 0$ because the complete three-node system is not strongly connected. Both single 3-cycles returned exactly the same positive value, $\varphi_s = 0.49356210998$.

All six systems nevertheless had identical ensemble predictive statistics: $H_2(Y|X) = 3h_2(0.75) \approx 2.433834373$ bits and $I(X;Y) = 3[1-h_2(0.75)] \approx 0.566165627$ bits. Under the homogeneous-state fully observed uniform-preference construction, $G_2$ has the same numerical value, $G_2 \approx 0.566165627$ bits for every topology. The equality of $I$ and $G_2$ here is specific to the symmetric construction and is not asserted as a general identity.

### D. Three-, four-, and five-cycle validation

For the single 3-cycle, the analytical expression $\varphi_s^{(3)}(\eta) = \min\{2\eta^3\log_2(2\eta), -3\log_2\eta\}$ was tested at 101 equally spaced values over $0.5 \leq \eta \leq 1$. The maximum discrepancy in $\varphi_s$ was $3.33\times10^{-16}$; corresponding cause/effect and intrinsic-differentiation quantities agreed at the same floating-point scale. The continuous optimum was $\eta_3^* \approx 0.827088719$ with $\varphi_{s,\max} \approx 0.821658013$ ibits.

For $n = 4$, selected reliability values spanning both limiting regimes and the predicted optimum reproduced the analytical result with maximum discrepancy $6.66\times10^{-16}$. The optimum was $\eta_4^* \approx 0.861148274$ with $\varphi_{s,\max} \approx 0.862665721$ ibits. Controls with cycle types 1+1+1+1, 2+2, and 3+1 all returned $\varphi_s = 0$.

For $n = 5$, the same tests gave maximum discrepancy $1.55\times10^{-15}$. At $\eta = 0.75$, PyPhi and the analytical expression both gave $\varphi_s \approx 0.277628687$ ibits. The predicted optimum $\eta_5^* \approx 0.884119884$ gave $\varphi_{s,\max} \approx 0.888430431$ ibits. Identity, 3+2, and 4+1 controls again returned zero whole-system integration.

### E. Independent directed-partition audit

The MIP derivation was tested independently of the full IIT repertoire calculation. All distinct directed cut matrices generated by the directed-set partition construction were enumerated for $n = 2, \ldots, 8$. For each matrix we recorded $k$, the number of actual cycle edges severed, and $M$, the number of possible directed connections severed. The audit considered 3, 22, 150, 1061, 7896, 61888, and 510313 distinct cut matrices for $n = 2, \ldots, 8$, respectively.

For every $n = 3, \ldots, 8$, the global minimum of $k/M$ had $k = 2$ and $M = \lfloor 4n^2/7 \rfloor$, exactly as proved in Appendix A. At $n = 2$, one- and two-edge cuts tied after normalization. Thus, the computational audit independently confirms the combinatorial MIP theorem rather than providing an empirical basis for it.

## X. DISCUSSION

### A. Local correspondence and global non-equivalence

The central result is not merely that EFE and integrated information can take different numerical values. The stronger statement is that the complete intact one-step predictive distribution can be identical while intrinsic causal integration differs. In the homogeneous state, $Q_{\sigma c}(Y|1^n) = Q_{\sigma m}(Y|1^n)$. Thus, the non-equivalence cannot be attributed simply to compressing a high-dimensional predictive distribution into a scalar free-energy value.

The distinction arises because IIT asks a counterfactual causal question. It evaluates intact, unconstrained, and causally partitioned repertoires and identifies the MIP. Two systems can therefore make the same prediction from their actual state while differing in what would happen under selective causal interventions. In compact form, predictive equivalence is not counterfactual causal equivalence.

This resolves an apparent tension in the original motivation of the study. The local algebraic connection is real: $i_{\text{diff}}$ is exactly a surprisal, while $i_{\text{spec}}$ and the partition-dependent cause/effect irreducibilities $\varphi_{c/e}$ are selectivity-weighted differences between surprisal coordinates. Under exact inference those coordinates can be represented as optimized VFE quantities. However, this local connection does not propagate to a global scalar identification because IIT retains additional intervention-sensitive information.

### B. Relation to previous IIT–FEP and active-inference work

The result clarifies the relation to previous attempts to connect IIT and the FEP. Olesen et al. [20] showed empirical covariation between surprisal and IIT-related quantities. Culture-based work [21], maximum-caliber proposals [22], and simulations of active-inference agents [23] likewise identify potentially meaningful relationships. Such covariance is entirely compatible with the present theorem. Correlation asks whether quantities tend to vary together over a sampled ensemble; functional identifiability asks whether one quantity determines the other. One pair with equal G and unequal $\varphi_s$ is sufficient to rule out a universal scalar map.

The broader consciousness-theory literature provides additional context. Safron's IWMT explicitly combines IIT, GNWT, the FEP, and active inference [19], while the Cogitate adversarial collaboration directly compared IIT and GNWT experimentally [25]. These broader frameworks remain relevant to theoretical synthesis, but GNWT does not supply a canonical scalar quantity directly analogous to $G$ or $\varphi_s$. For that reason, the present mathematical theorem deliberately focuses on the two scalar constructions for which a precise identifiability question can be posed.

### C. Role of the 2026 intrinsic-differentiation refinement

The 2026 intrinsic-differentiation constraint [6] qualitatively changes the reliability dependence. Partition-based irreducibility alone increases toward the deterministic limit, but intrinsic differentiation $D_n = -n\log_2\eta$ vanishes as $\eta \to 1$. Their competition makes $\varphi_s$ zero at both complete randomness and complete determinism and produces a unique interior maximum. In large cycles, each local interaction becomes nearly deterministic at the optimum while the joint specified-state probability approaches one half, leaving an asymptotic maximum of one ibit. This behavior cannot be inferred from the pre-2026 partition term alone.

### D. Preference dependence and what the theorem does not claim

The optimization mismatch shown for uniform future-state preferences is illustrative rather than universal. EFE is defined relative to a generative model and prior preferences, and Eq. (54) shows explicitly that its optimum can be shifted by changing those preferences. Revised IIT $\varphi_s$, by contrast, is intended to characterize intrinsic cause–effect power without an externally supplied preference distribution. Agreement between their extrema is therefore not generically expected, but it can be imposed by choosing matching preferences.

Likewise, the theorem does not imply that active inference is generally insensitive to topology. A richer generative model can explicitly encode architecture-dependent hidden states, transition parameters, interventions, or policies. In such a construction, topology can affect EFE. The claim is one of non-sufficiency: $G$ alone, and in the present counterexample even the intact one-step predictive distribution, is not enough to reconstruct revised IIT system integration.

### E. Physical and conceptual interpretation

The result complements statistical-physical approaches relating integration to information geometry and thermodynamic irreversibility [11,12,17,18]. Such connections are valuable because they identify shared mathematical structures, but a scalar divergence or thermodynamic cost does not by itself specify which causal edges are present or how the system changes under interventions. The IIT partition construction retains precisely this missing organization.

A successful bridge between FEP/active-inference and IIT descriptions would therefore need additional topology-sensitive or intervention-sensitive variables. Schematically, one would require a relation more like $\varphi_s = F(G,\mathcal{T},s,\ldots)$ where $\mathcal{T}$ represents causal structure, rather than $\varphi_s = f(G)$ alone. The local free-energy-coordinate identities derived in Sec. III can form part of such a bridge, but they are insufficient by themselves.

### F. Limitations and open directions

The noisy permutation channels are deliberately minimal analytically tractable systems, not realistic neural models. The homogeneous state is chosen specifically to provide a permutation-invariant predictive counterexample; generic states need not be predictively indistinguishable. The analysis is one-step and focuses on system $\varphi_s$ rather than the complete IIT structure of distinctions and relations. Multi-step policies, learning, hierarchical inference, action-dependent environments, and full cause–effect structures may introduce additional relationships.

The theorem also does not establish that either IIT or the FEP/active-inference framework is superior as a theory of consciousness or cognition. Nor does it adjudicate broader empirical debates among IIT, GNWT, and other consciousness theories [25,26,29–31]. It establishes a narrower mathematical constraint: scalar expected free energy is not a sufficient statistic for revised IIT system integrated information over the noisy-permutation family.

## XI. CONCLUSIONS

This work began with the possibility of a mathematical connection between IIT and free-energy descriptions and identifies both where that connection succeeds and where it fails. At the local level, revised IIT quantities admit exact surprisal coordinates: intrinsic differentiation is itself a state-specific surprisal, while intrinsic specification and partition irreducibility are selectivity-weighted differences between intact, unconstrained, and partitioned surprisals. Under exact inference these can be represented as optimized variational-free-energy coordinates.

At the system level, however, the correspondence does not imply scalar equivalence. Noisy permutation networks provide an explicit family in which distinct causal topologies have exactly the same intact one-step predictive distribution and hence the same state-risk expected free energy under a common generative model, while revised IIT assigns different whole-system integration. A multicycle permutation and a single-cycle permutation can satisfy $G_1 = G_2$ but $\varphi_{s,1} \neq \varphi_{s,2}$. Therefore no universal scalar mapping $\varphi_s = f(G)$ exists over the family.

For a noisy single $n$-cycle we obtained the exact expression $\varphi_s^{(n)}(\eta) = \min\{2\eta^n\log_2(2\eta), -n\log_2\eta\}$, proved the two-edge MIP structure, and derived the unique optimal reliability and large-system limits $\eta_n^* \to 1$, $(\eta_n^*)^n \to 1/2$, and $\varphi_{s,\max}^{(n)} \to 1$ ibit. The fraction of permutation architectures that are single cycles is only $1/n$, so the proportion of predictively equivalent architectures lacking whole-system integration approaches unity with increasing $n$.

The central distinction is therefore between prediction and counterfactual causal organization. Any formal synthesis of active inference and IIT must preserve the latter rather than attempting to recover intrinsic causal integration from expected free energy alone.

## ACKNOWLEDGMENTS

This study was financially supported, in part, by the grant number KSAC2025-14, the Japan Science and Technology Agency.

## DATA AVAILABILITY

The source code and numerical data supporting this study are openly available [37].

## APPENDIX A: MINIMUM-INFORMATION PARTITION OF A DIRECTED CYCLE

This appendix gives the full combinatorial proof underlying Eq. (41). Consider the directed $n$-cycle $v_1 \to v_2 \to \cdots \to v_n \to v_1$. Let $\theta$ be an admissible IIT 4.0 directed-set system partition. Each partition block is assigned one of three causal directions: CAUSE ($C$), EFFECT ($E$), or BIDIRECTIONAL ($B$). Let $k(\theta)$ denote the number of actual causal edges of the cycle severed by $\theta$, and let $M(\theta)$ denote the number of possible directed connections severed by the corresponding cut matrix. Because Eq. (39) gives $\varphi_{c/e}(\theta) = k\eta^n\log_2(2\eta)$, the normalized MIP is determined entirely by $R(\theta) = k/M$. For fixed $k$, let $R_k$ denote the minimum of $R(\theta)$ over all partitions that sever $k$ actual cycle edges.

### 1. Uncut-path structure

For two distinct partition blocks, an ordered inter-block connection remains uncut only when its source lies in a $C$ block and its target lies in an $E$ block, i.e., only for $C \to E$. Connections within the same block remain uncut. Removing $k$ actual cycle edges decomposes the directed cycle into exactly $k$ paths of uncut actual edges. Along one such path, an inter-block boundary can be crossed without severing only as $C \to E$. Once the path enters an $E$ block, it cannot cross another inter-block boundary while remaining uncut. Therefore, each uncut path contains at most two blocks, and if it contains two their order is $C \to E$.

A partition block need not initially be contiguous on the cycle. However, if one block intersects more than one uncut path, splitting that block into separate blocks path by path leaves every actual cycle edge with the same cut/uncut status and hence leaves $k$ unchanged. Connections between the newly separated blocks may become cut, so $M$ cannot decrease. Therefore, for fixed $k$, an $M$-maximizing partition can be chosen with every block contained within a single uncut path. Likewise, a one-block uncut path can be labeled $B$ without decreasing $M$, because relabeling it $B$ cannot create new uncut inter-block connections. Thus, an $M$-maximizing fixed-$k$ partition may be taken to consist of $q$ two-block $C \to E$ paths and $k-q$ single $B$ blocks.

### 2. Bound for fixed $k$

Let the $q$ $C$-block sizes be $c_1, \ldots, c_q$, the $q$ $E$-block sizes $e_1, \ldots, e_q$, and the $k-q$ B-block sizes $b_1, \ldots, b_{k-q}$. Let $C$, $E$, and $B$ denote their respective total numbers of nodes, with $C+E+B = n$. If $U = n^2-M$ is the number of possible ordered connections left uncut, then within-block connections and all $C\to E$ connections give

$$U = \Sigma c_j^2 + \Sigma e_j^2 + \Sigma b_j^2 + CE. \tag{A1}$$

By the Cauchy–Schwarz inequality,

$$U \geq C^2/q + E^2/q + B^2/(k-q) + CE. \tag{A2}$$

Minimizing the right-hand side subject to $C+E+B = n$ gives, for $q \geq 1$,

$$U \geq n^2(q+2)/[k(q+2)+q(2-q)]. \tag{A3}$$

Define $g_k(q) = (q+2)/[k(q+2)+q(2-q)]$. Direct subtraction gives

$$g_k(q)-g_k(1) = (q-1)(3q-2) /\{(3k+1)[k(q+2)+q(2-q)]\}. \tag{A4}$$

For admissible $q > 1$ the numerator and denominator are positive, so $g_k(q) > g_k(1)$. Thus, the smallest lower bound on U, and hence the largest possible M, occurs at $q = 1$. The $q = 0$ case gives the weaker bound $U \geq n^2/k$. Therefore, every $k$-edge partition satisfies

$$U \geq 3n^2/(3k+1), \quad M \leq n^2(3k-2)/(3k+1). \tag{A5}$$

Consequently,

$$k/M \geq k(3k+1)/[(3k-2)n^2]. \tag{A6}$$

### 3. Exact optimum for $k = 2$

For $k = 2$ the optimal structure contains one two-block $C \to E$ path and one $B$ block. Let their sizes be

$a$, $c$, and $b$, with $a+b+c = n$. The uncut possible connections are

$$U = a^2+b^2+c^2+ac. \tag{A7}$$

Using $n = a+b+c$,

$$7U-3n^2 = (7/4)(a-c)^2 + (1/4)[3(a+c)-4b]^2 \geq 0. \tag{A8}$$

Hence $U \geq 3n^2/7$ and, because $U$ is an integer, $U \geq \lceil 3n^2/7 \rceil$. Therefore

$$M \leq \lfloor 4n^2/7 \rfloor. \tag{A9}$$

The continuous optimum has the block-size ratio $a$:$b$:$c$ = 2:3:2. The integer bound is attainable for every $n \geq 3$. Write $n = 7m+r$, $r \in \{0, \ldots, 6\}$, and choose the integer block sizes listed in Table II. Direct substitution into Eq. (A7) gives $U = \lceil 3n^2/7 \rceil$ in every residue class.

**TABLE II. Explicit integer block sizes attaining $U = \lceil 3n^2/7 \rceil$ for $n = 7m+r$.**

| $r$ | $a$ | $b$ | $c$ |
|---|---|---|---|
| 0 | $2m$ | $3m$ | $2m$ |
| 1 | $2m$ | $3m+1$ | $2m$ |
| 2 | $2m$ | $3m+1$ | $2m+1$ |
| 3 | $2m+1$ | $3m+1$ | $2m+1$ |
| 4 | $2m+1$ | $3m+2$ | $2m+1$ |
| 5 | $2m+1$ | $3m+2$ | $2m+2$ |
| 6 | $2m+2$ | $3m+2$ | $2m+2$ |

Thus the exact *fixed*-$k$ = 2 optimum is

$$M_2^{\max} = \lfloor 4n^2/7 \rfloor, \quad R_2 = 2/\lfloor 4n^2/7 \rfloor. \tag{A10}$$

### 4. Global comparison with $k = 1$ and $k \geq 3$

For $k = 1$, Eq. (A6) gives $R_1 \geq 4/n^2$. For $k \geq 3$, define $f(k) = k(3k+1)/(3k-2)$. Its derivative $f'(k) = [9k^2-12k-2]/(3k-2)^2$ is positive for $k \geq 2$, so

$$R_k \geq 30/(7n^2) \text{ for } k \geq 3. \tag{A11}$$

For $n = 3$, $R_2 = 2/5 = 0.4$, which is below both 4/9 and 30/63. For $n \geq 4$, $\lfloor 4n^2/7 \rfloor > n^2/2$ and $\lfloor 4n^2/7 \rfloor > 7n^2/15$, so $R_2$ is strictly below both the $k = 1$ and $k \geq 3$ lower bounds. Hence the global MIP has $k_{\mathrm{MIP}} = 2$ for every $n \geq 3$. For $n = 2$, one-edge and two-edge partitions tie after normalization, $k/M = 1$, and the IIT tie-resolution procedure selects the two-edge partition because its unnormalized integrated information is larger. Equation (42) therefore extends to $n = 2$. An independent exhaustive computational cross-check is reported in Sec. IX E.